\documentclass[12pt]{article}
\usepackage{graphicx}
\usepackage{hyperref}
\usepackage{amsmath}

\newcommand\pubnumber{ IRMP-CP3-24-02}
\newcommand\GeV{\text{ GeV}}
\newcommand {\abs}[1]{\left\lvert#1\right\rvert}
\newcommand\pubdate{\today}

\newcommand{\be}{\begin{equation}}
\newcommand{\ee}{\end{equation}}

\def\institute{
${}^{1}${Centre for Cosmology, Particle Physics and Phenomenology (CP3),
Universit\'{e} Catholique de Louvain, B-1348 Louvain-la-Neuve, Belgium}\\~
}

\def\authemails{\footnote{simone.tentori@uclouvain.be}}

\def\Title#1{\begin{center} {\Large #1 } \end{center}}
\def\Author#1{\begin{center}{ \sc #1} \end{center}}
\def\Address#1{\begin{center}{ \it #1} \end{center}}

\newcommand\pubblock{\rightline{\begin{tabular}{l} \pubnumber\\
         \pubdate  \end{tabular}}}
\newenvironment{Abstract}{\begin{quotation}  }{\end{quotation}}
\newenvironment{Presented}{\begin{quotation} \begin{center} 
             PRESENTED AT\end{center}\bigskip 
      \begin{center}\begin{large}}{\end{large}\end{center} \end{quotation}}
\def\Acknowledgements{\bigskip  \bigskip \begin{center} \begin{large}
             \bf ACKNOWLEDGEMENTS \end{large}\end{center}}

\def\beq{\begin{equation}}
\def\eeq#1{\label{#1}\end{equation}}
\def\eeqn{\end{equation}}

\def\beqa{\begin{eqnarray}}
\def\eeqa#1{\label{#1}\end{eqnarray}}
\def\eeqan{\end{eqnarray}}

\let\bar=\overbar

\def\Dslash{\not{\hbox{\kern-4pt $D$}}}
\def\dslash{\not{\hbox{\kern-2pt $\del$}}}

\def\ee{e^+e^-}

\def\msb{{\bar{\ssstyle M \kern -1pt S}}}

\begin{document}
\begin{titlepage}
\pubblock

\vfill
\Title{Top-philic ALP phenomenology at the LHC}
\vfill
\Author{
{Simone Tentori}\authemails}
\Address{\institute}
\vfill
\begin{Abstract}
{
I present an exploration of the LHC phenomenology of the so-called top-philic ALP in the mass range between tens and hundreds of GeV. 
Searches through resonant production, such as ALP production in association with a $t\bar t$ pair are shown to be complementary to precision measurements of $t \bar t$  and $t\bar t t \bar t$ final states.  The case where the ALP decays invisibly is also considered looking at final states with missing energy signatures.
}
\end{Abstract}
\vfill
\begin{Presented}
$16^\mathrm{th}$ International Workshop on Top Quark Physics\\
(Top2023), 24--29 September, 2023
\end{Presented}
\vfill
\end{titlepage}
\def\thefootnote{\fnsymbol{footnote}}
\setcounter{footnote}{0}

\section{Introduction}
Although axion was originally proposed to solve the strong CP problem, nowadays axion-like particles (ALP) are a common feature of many BSM scenarios.
The ALP is a pseudoscalar singlet $(a)$ interacting with the SM particles with an approximate shift-symmetric interaction $a\rightarrow a+c$, where $c$ is a constant. The shift-symmetry is originally due to the fact that $a$ is a pNGB of a spontaneously broken symmetry. As such,
the ALP is naturally light compared to other BSM states.

In this presentation we focus on a ALP that only couples with the top-quark at the leading order:
\begin{equation}
\mathcal{L}_{\rm int}:=c_t\frac{\partial^\mu a}{f_a}\bar t_{\rm R}\gamma_\mu t_{\rm R}.
\end{equation}
This particular ALP set-up has gained a lot of interest recently, being the subject of many recent studies~\cite{Blasi:2023hvb,Bonilla:2021ufe,Carmona:2022jid,Phan:2023dqw,Rygaard:2023dlx,Bruggisser:2023npd}. 
Flavour physics\cite{Bauer:2021mvw} and astrophysical \cite{Chala:2020wvs} probes effectively bound the ALP couplings when the ALP mass is small ($m_a<10\GeV$) while resonant searches in the $t\bar t$ spectrum strictly constrain the region $m_a>2m_t$. The intermediate region $10\GeV<m_a<200\GeV$ is the most difficult to constrain and is the main topic of this study. In Sec. \ref{sec:ALPgeneral} the  top-philic ALP couplings to SM together with the production channels and decays rates are analyzed, including the bounds obtainable from already existing experimental searches. Sec. \ref{sec:newprobes} is dedicated to new probes for the top-philic ALP analyzing both the case in which it decays visibly to SM particles, and the case in which it decays invisibly, acting as dark matter (DM) portal. These proceedings present results  published in Ref.\cite{Blasi:2023hvb}, in the main paper are listed all the experimental measurements used for setting bounds.
\section{Top-philic ALP at colliders}\label{sec:ALPgeneral}
\subsection{Top-philic ALP interactions}
\begin{figure}
    \centering
    \includegraphics[width=0.57\textwidth]{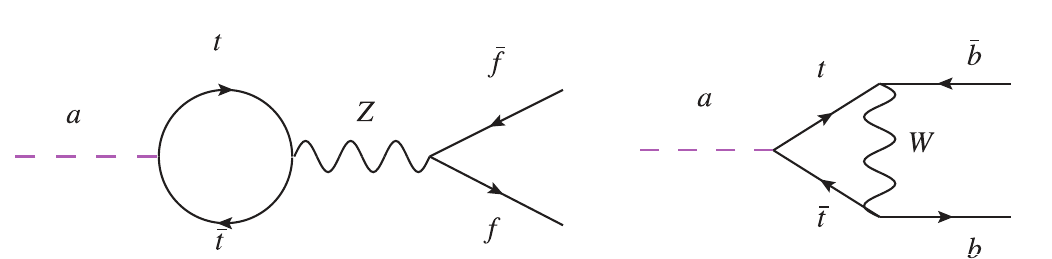}
    \caption{Loop diagrams contributing to the running of $c_{f\neq t}$ couplings.}
    \label{fig:loops}
\end{figure}

Starting from a specific UV model\cite{Blasi:2023hvb,Esser:2023fdo} it is possible to obtain an ALP coupling only with top-quark at tree-level
\begin{equation}
\label{eq:top-philic}
\mathcal{L}_{\text{top-philic}} = \frac{1}{2} (\partial_\mu a)^2 -\frac{1}{2} m_a^2 a^2 + c_t \frac{\partial^{\mu} a}{f_a} \bar t_R \gamma_{\mu} t_R\, .
\end{equation}
This nice feature is lost at the loop-level, where diagrams such as those in Fig. \ref{fig:loops} generate effective couplings to all SM fermions.
These couplings play an important role despite being loop-generated, as will be shown later.
The modifications to the tree-level couplings are logarithmically dependent on the cut-off scale of the EFT $\Lambda$, the divergent part has been computed~\cite{Bonilla:2021ufe,Bauer:2020jbp} giving\begin{equation}
\label{eq:cthird}
c_t(m_t) = c_t(\Lambda) \left( 1 - 9 \frac{y_t^2}{16\pi^2} \log \frac{\Lambda}{m_t} \right), \quad
c_b(m_t) = 5 c_t(\Lambda) \frac{y_t^2}{16 \pi^2} \log \frac{\Lambda}{m_t}\,,
\end{equation}
while for the other fermions $f=u,d,c,s,e,\mu,\tau$ one has
\begin{equation}
\label{eq:clight}
c_f(m_t) = - 12\, c_t(\Lambda) \frac{y_t^2}{16 \pi^2} T_3^f \log \frac{\Lambda}{m_t}\,,
\end{equation}
where $T_3^f$ represents the isospin component of $f_L$.
\subsection{Top-philic ALP production and decays}\label{sec:proddec}
\begin{figure}[htb]
    \centering
    \includegraphics[width=0.54\textwidth]{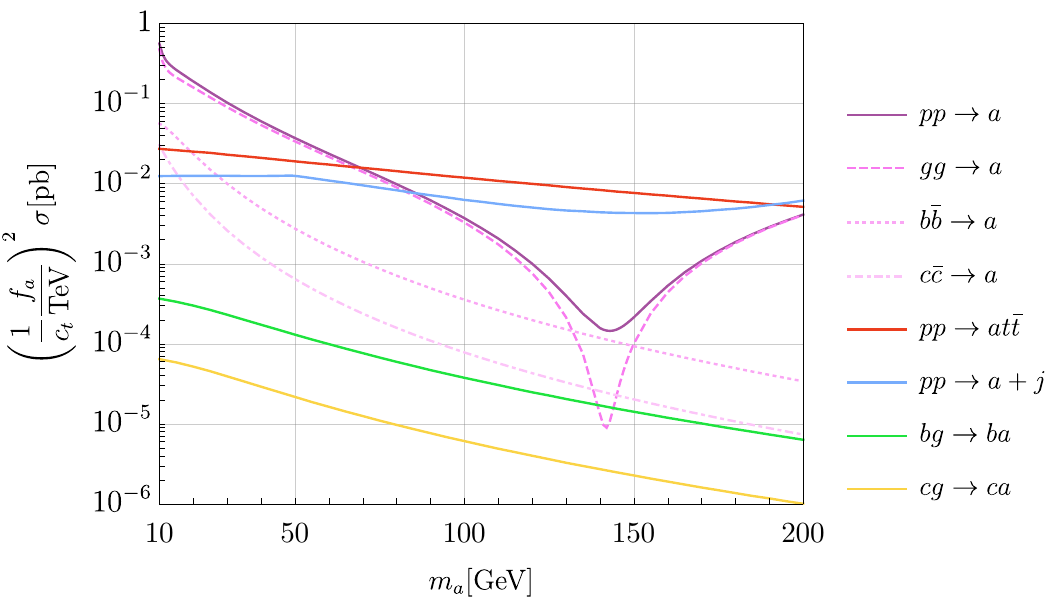}
\includegraphics[width=0.44\textwidth]{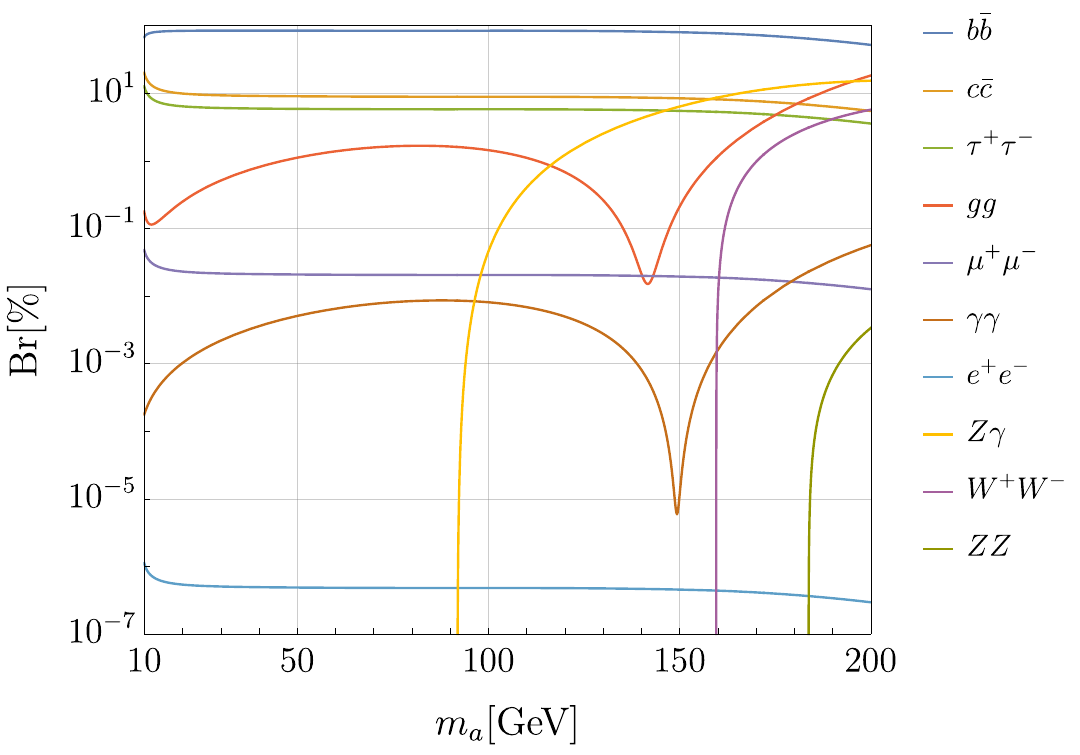}
    \caption{\textbf{Left:} top-philic ALP production at LHC. \textbf{Right:} top-philic ALP decay rates into SM particles.}
    \label{fig:proddecay}
\end{figure}
The production cross sections for the top-philic ALP at LHC are presented in Fig. \ref{fig:proddecay} left.
The most important production modes are the ALP production in association with a $t\bar t$ pair (red) and the ALP production in association with one-jet (indigo).
The gluon fusion (purple), in the top-philic ALP case, is very suppressed w.r.t. the case of a generic ALP, due to the absence of a tree-level contact term with the gluons in \eqref{eq:top-philic}.
The gluon fusion is a typical example in which one should carefully consider the contribution to the process coming from other fermions, even if effectively two-loops. In fact in the considered mass range ($m_a\ll2m_t$ and $m_a>m_{q\neq t}$) the following expansion for the gluon fusion is valid
\begin{align}
\label{eq:Cgg2loopintermediate}
\sigma(gg\rightarrow a)\propto{\Big|}_{2m_{b}\ll m_{a} \ll 2m_{t}} &\simeq \abs{ -\frac{m_a^2}{24 m_t^2} c_t(m_{t}) +\frac{1}{2}  c_b(m_{t}) }^2.
\end{align}
First of all, it is worth noticing that only the $t$-quark and $b$-quark contributions remain, the other light quarks cancel each other being the effective coupling proportional to the isospin (as for Eq. \ref{eq:clight}). On the other hand, the contribution coming from the $b$-quark is numerically much larger than the one coming from the $t$-quark for small $m_a$. The contribution coming from the $t$-quark increases with $m_a$, this effect together with the opposite sign between the $b$-quark and $t$-quark contributions gives the strange dip-shaped line for $gg\rightarrow a$ in Fig. \ref{fig:proddecay}.
    Given the importance of the induced $b$-quark coupling, a precise computation of this process would require a two-loop calculation. However, the overall magnitude of $gg\rightarrow a$ (even far from the dip) and the small branching ratio into photons (Fig. \ref{fig:proddecay}) together make this channel ineffective in constraining the top-philic ALP coupling $c_t/f_a$. The conclusions extracted here therefore do not depend on the precise computation (at the two-loop level) of this process.
\subsection{LHC constraints from existing BSM searches}
\begin{figure}[htbp]
    \centering
    \includegraphics[width=0.8\textwidth]{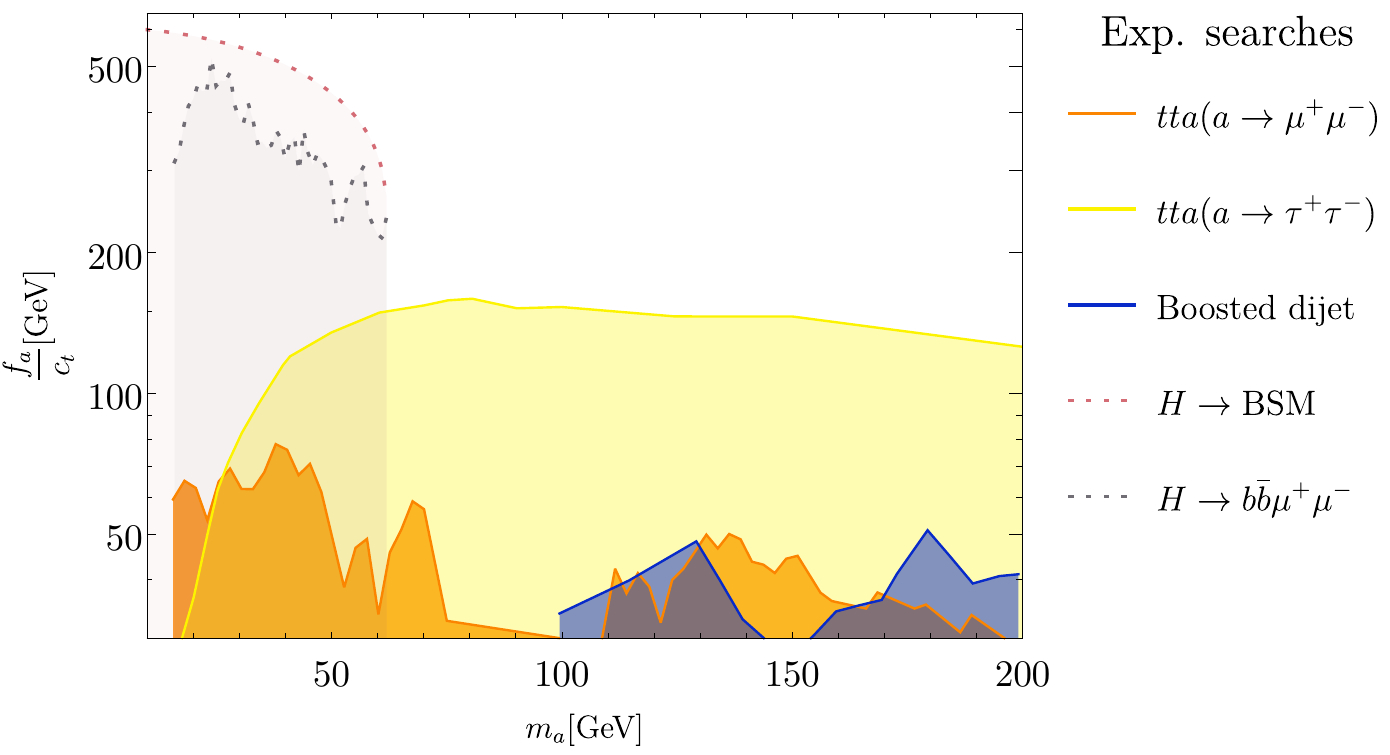}
    \caption{   Bounds on ${f_a}/{c_t}$ obtained using LHC constraints from 
   existing searches targeting BSM states, which can be re-interpreted for the top-philic ALP.}
    \label{fig:Expprevious}
\end{figure}
The most stringent experimental search, given the relatively large $t\bar t a$ cross-section is the one targeting the $t\bar t$ production in association with two opposite sign leptons $\tau^+\tau^-$ (yellow line) or $\mu^+\mu^-$ (orange line) in Fig. \ref{fig:Expprevious}, coming from the decay of the top-philic ALP. The $gg\rightarrow a$ channel allows searches in $b\bar b$ or $\gamma \gamma$ resonances, as anticipated in the previous section the bound obtained is so small that it is not even present in Fig. \ref{fig:Expprevious}. The $a+j$ final state gives a signature for boosted dijet searches, once considering $a$ decaying into 2 $b$-quarks (blue line).

On a general ground, the Higgs decay can be used to put bounds over $c_t/f_a$, these lines are however dotted in Fig. \ref{fig:Expprevious} due to the possible presence of a dim-6 operator 
\begin{align}
    \mathcal{L}^{(6)}= \frac{c^{(6)}_{a H}}{f_{a}^{2}}\phi^\dagger\phi\, \partial_\mu a \partial^\mu a\equiv  \frac{c^{(6)}_{a H}}{f_{a}^{2}} \mathcal{O}^{(6)}_{a H}\,,
\end{align}
that would significantly affect this bound.
\section{New probes for the top-philic ALP}\label{sec:newprobes}
\begin{figure}[htb]
    \centering
    \includegraphics[width=0.8\textwidth]{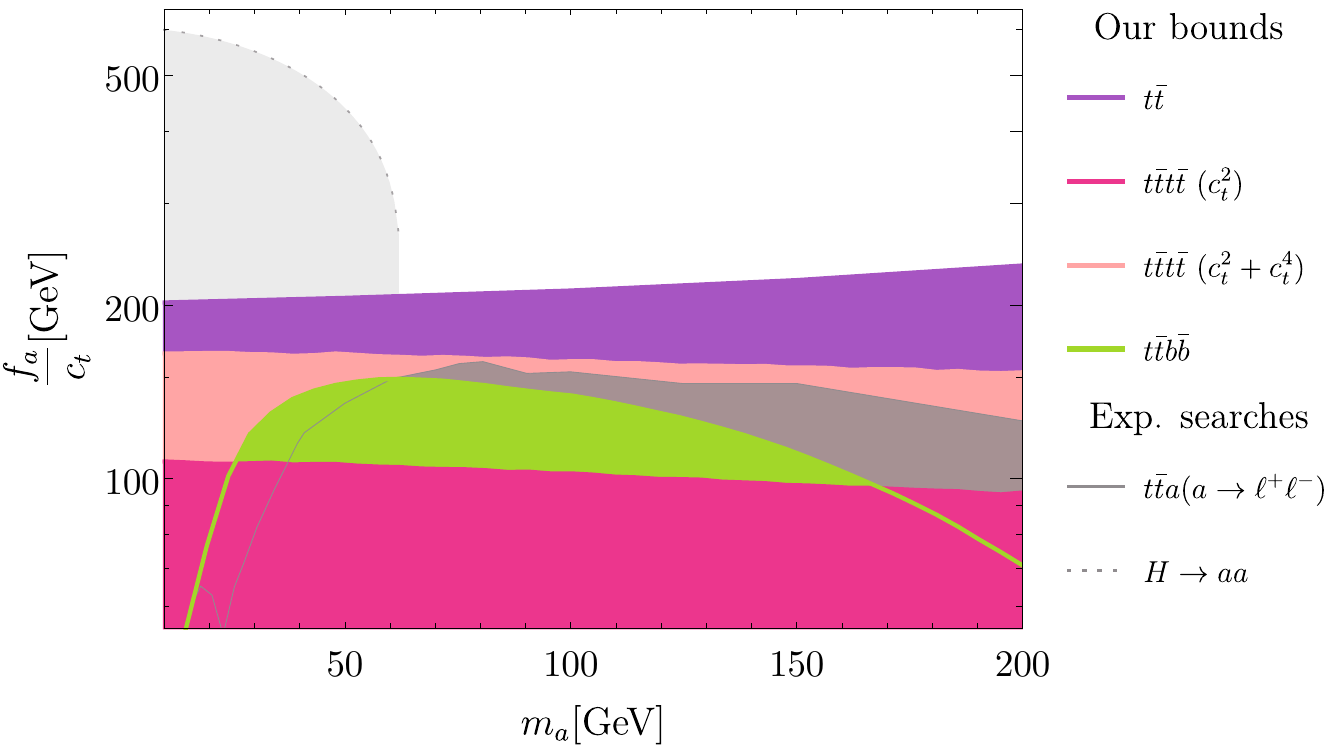}     \caption{   
Summary plot of exclusion limits obtained for the visible ALP case. In the figure are depicted both the new probes and the best experimental existing bounds.}
 \label{fig:finalvis}

    \end{figure}

\begin{figure}
    \centering
        \includegraphics[width=0.8\textwidth]{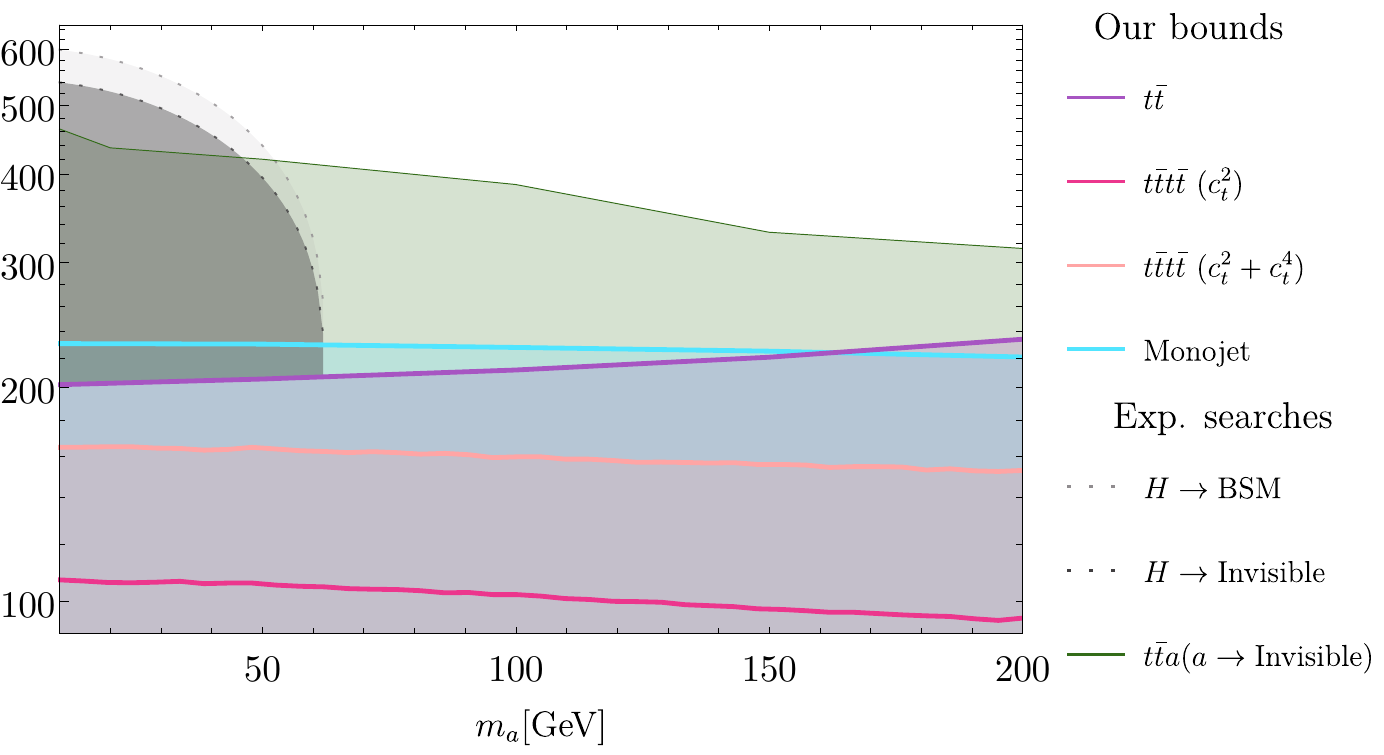}
    \caption{Combination of exclusions from current experimental searches and the proposed probes for a top-philic ALP decaying invisibly with a $100\%$ BR.}
    \label{fig:finalinv}
\end{figure}

This section is dedicated to new probes for the top-philic ALP and the related new bounds obtainable for $c_t/f_a$ . In particular, the following measurements are considered:
\begin{itemize}
    \item $t\bar t t \bar t$ total cross section measurement;
    \item $t\bar t$ differential distributions;
    \item $t\bar tb\bar b$ total cross-section;
    \item Monojet final state plus missing energy.
\end{itemize}
In the first two cases the ALP is off-shell and the bounds obtained are appliable both for an ALP decaying visibly, both for one decaying invisibly.
The third case is a resonance search, and it is valid for an ALP decaying into SM particles. The last bound is valid only if the ALP decays invisibly giving a final missing energy signature, this is the scenario of a top-philic ALP being a portal to DM. The final exclusion plots for $c_t/f_a$ are in Fig. \ref{fig:finalvis} for the visible case and in Fig. \ref{fig:finalinv} for the invisible one.
    
\subsection{Off-shell ALP}
\paragraph{$\mathbf{t\bar t t\bar t}$ total cross-section:} In the considered mass range, top-philic ALPs contribute non-resonantly to four-top production at hadron colliders. This can happen in many different ways, through $t$- or $s$-channel-like diagrams involving one or even two ALPs, via $gg$ or $q\bar{q}$ initiated processes.  In Fig. \ref{fig:finalvis} are shown the $95\%$ C.L. bounds obtained: the fuchsia line is the bound obtained considering only top-philic ALP diagram interfering with SM (order $c_t^2$) while the pink one is obtained considering also the top-philic ALP diagrams squared ($c_t^4$). The bound obtained for the last case is around $150\GeV$ for $f_a/c_t$ being better than the direct experimental searches in $t\bar t\ell^+\ell^-$ final state.
\paragraph{$\mathbf{t\bar t}$ virtual corrections:} many recent studies have focused on the corrections to $t\bar t$ distributions given by the possible existence of a top-philic ALP ~\cite{Blasi:2023hvb,Phan:2023dqw,Bruggisser:2023npd,Esser:2023fdo}. The corrections to the SM cross section at order $c_t^2$ can be separated in virtual corrections (due to loops involving the top-philic ALP) and real emission of $a$.
Even if the former dominate, the phenomenological importance of the real emission diagrams depends on whether a cut is imposed on $p_{\rm T}(a)$  \cite{Blasi:2023hvb} or the full $a$ phase-space is considered \cite{Phan:2023dqw}.  The bound presented here is obtained by trying different combinations of statistically independent  $t\bar t$  measurements in $m(t\bar t)$ and $p_{\rm T}(t)$ and selecting the most stringent one. The final bound obtained is around $200\GeV$ for $f_a/c_t$ being the best for the visible top-philic ALP.
\subsection{On-shell top-philic ALP}
\paragraph{$\mathbf{t\bar t b\bar{b}}$ total cross section:} the green line in Fig. \ref{fig:finalvis} is the $95\%$ C.L. bound obtained using total cross-section measurements of $t\bar t b \bar b$.  The simulated signal events are subjected to a minimal set of parton-level cuts associated to a final state with two $b$-jets  in the LHC experiments. At this moment there are no experimental measurements for $b\bar b$ resonances produced in association with  $t\bar t$, a dedicated search could be then really useful to put better bounds or even as a possible discovery channel.
\paragraph{Invisible top-philic ALP:} The top-philic ALP can play the role of a portal to DM. If $m_{\rm DM}>m_b$, then the ALP will decay preferably invisibly. In this case, missing energy signatures at collider can be exploited, as in $t\bar t$ plus missing energy (dark-green) ~\cite{ATLAS:2021hza} or in final QCD jet plus missing energy (light-blue) ~\cite{ATLAS:2021kxv} in Fig. \ref{fig:finalinv}.
\section{Conclusion}
I have illustrated the main characteristics of the top-philic ALP  such as its coupling to SM fermions, the production and decay channels. Existing experimental measurements have been proven to fall short in effectively constraining the top-philic ALP coupling $c_t/f_a$.  This is the reason why many recent works have focused on combining different strategies to observe or exclude a top-philic ALP. We have shown that indirect detection through $t$-quark final states appear to be very promising.

\Acknowledgements
ST is supported by a FRIA Grant of the Belgian Fund for Research, F.R.S.-FNRS. This work has been realized in collaboration with S. Blasi, F. Maltoni, A. Mariotti, K. Mimasu and D. Pagani.
\newpage

\bibliography{eprint}{}
\bibliographystyle{unsrt}
 
\end{document}